\documentclass[11pt,letterpaper]{article}%
\usepackage{amsmath}
\usepackage{epsfig}
\usepackage{graphicx}%
\usepackage{amsfonts}%
\usepackage{amssymb}

\begin{document}
%
\begin{titlepage}
\begin{flushright}
HEP-PH/0408XXX
\end{flushright}
\vspace*{28mm}
\begin{center}
\huge{Recent progress on the rare decay $K_{L}\rightarrow\pi^{0}\mu^{+}\mu
^{-}$}
\end{center}
\vspace*{8mm}
\begin{center}
\Large{Christopher Smith\footnote{Christopher.Smith@lnf.infn.it}}
\end{center}
\vspace*{3mm}
\begin{center}
INFN - Laboratori Nazionali di Frascati, I-00044 Frascati, Italy
\end{center}
\vspace*{4mm}
\begin{center}
August 1, 2004
\end{center}
\vspace*{7mm}
\begin{abstract}
The rare decay  $K_{L}\rightarrow\pi^{0}\mu^{+}\mu^{-}%
$ has a significant CP-conserving
contribution. The reliable theoretical estimation of this piece from the experimental
$K_{L}\rightarrow\pi^{0}\gamma\gamma
$ branching ratio is shortly reviewed. Then,
we discuss the exceptional sensitivity of the combined set of decays, into
$\pi^{0}\nu\bar{\nu}$, $\pi^{0}e^{+}e^{-}$ and $\pi^{0}\mu^{+}\mu^{-}$,
to New Physics signal, and also, interestingly, to New Physics nature.
\end{abstract}
\vspace*{10mm}
\begin{center}
\it{Short talk given at Da$\Phi$ne 2004: Physics at meson factories, \\
June 7 - 11, 2004, Frascati, Italy. \\
http://www.lnf.infn.it/conference/2004/dafne04}
\end{center}
\vspace*{10mm}
\end{titlepage}%

\newpage

\section{Introduction}

Studies of direct CP-violation are important to test the Standard Model, and
possibly to discover New Physics effects. Here, we consider the following rare
$K_{L}$ modes%
\[%
\begin{tabular}
[c]{lccc}
& DCPV & ICPV & CPC\\\hline
$K_{L}\rightarrow\pi^{0}\nu\bar{\nu}$ & $\sim100\%$ & $\sim0\%$ & $\sim0\%$\\
$K_{L}\rightarrow\pi^{0}e^{+}e^{-}$ & $\sim40\%$ & $\sim60\%$ & $\sim0\%$\\
$K_{L}\rightarrow\pi^{0}\mu^{+}\mu^{-}$ & $\sim30\%$ & $\sim35\%$ & $\sim
35\%$\\\hline
\end{tabular}
\]
The direct CP-violating (DCPV), indirect CP-violating (ICPV) and CP-conserving
(CPC) contributions originate from the processes depicted on Fig.1. While the
theoretical complexity increases for $\ell^{+}\ell^{-}$ modes, recent
experimental results for $K_{S}\rightarrow\pi^{0}\ell^{+}\ell^{-}%
$\cite{NA48as} and $K_{L}\rightarrow\pi^{0}\gamma\gamma$%
\cite{KTeV_KLpgg,NA48_KLpgg} now permit reliable theoretical estimates for
ICPV and CPC, making them competitive with the $\nu\bar{\nu}$
one.\begin{figure}[th]
\begin{center}%
\[
\includegraphics[width=13cm]{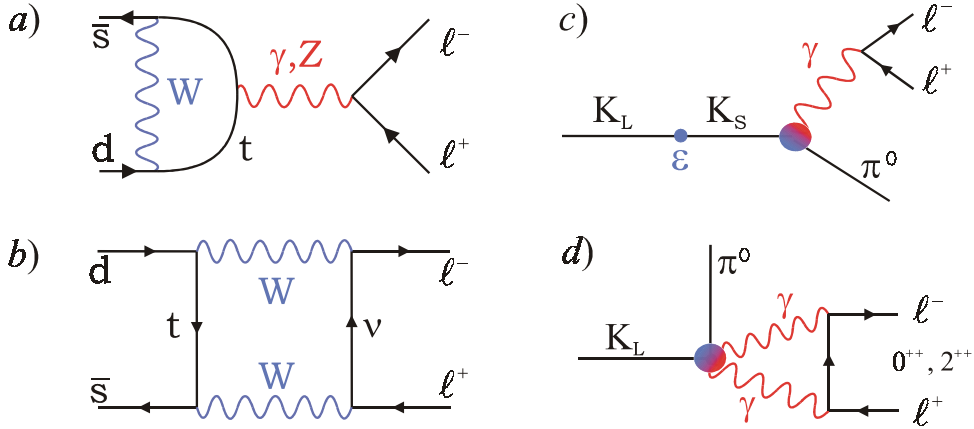}  \hspace{1cm}%
\]
\end{center}
\caption{DCPV (a,b), ICPV (c) and CPC (d) contributions.}%
\label{fig1}%
\end{figure}

The CPC contribution, as shown on Fig.1d, proceeds through two photons, which
can be in a scalar $0^{++}$ or tensor $2^{++}$ state. The former is helicity
suppressed, hence contribute only for $\mu^{+}\mu^{-}$, while the later, much
smaller, is the dominant one for $e^{+}e^{-}$ (see Ref.\cite{BDI}). Our work
was to estimate the $0^{++}$ CPC contribution\cite{OurWork}.

\section{CP--conserving contribution}

At leading order in Chiral Perturbation Theory (i.e. $\mathcal{O}\left(
p^{4}\right)  $ here), the process proceeds through a charged meson loop
followed by a photon loop, $K_{L}\rightarrow\pi^{0}P^{+}P^{-}$, $P^{+}%
P^{-}\rightarrow\gamma\gamma\rightarrow\ell^{+}\ell^{-}$ with $P=\pi,K$, see
Fig.2a. In addition, it is factorized and $\left(  P^{+}P^{-}\right)
_{0^{++}}\rightarrow\gamma\gamma\rightarrow\ell^{+}\ell^{-}$ can be computed
separately. This subprocess is then strictly similar to $K_{S}\rightarrow
\left(  \pi^{+}\pi^{-}\right)  _{0^{++}}\rightarrow\gamma\gamma\rightarrow
\ell^{+}\ell^{-}$ \cite{OurWork,EP}.

This factorization holds when one can parametrize the vertex $\mathcal{M}%
\left(  K_{L}\rightarrow\pi^{0}P^{+}P^{-}\right)  =G_{8}m_{K}^{2}a^{P}\left(
z\right)  $ (see \cite{OurWork,CEP}), with $z=\left(  p_{P^{+}}+p_{P^{-}%
}\right)  ^{2}/m_{K}^{2}$ the $P^{+}P^{-}$ invariant mass and $a^{P}\left(
z\right)  $ some function ($a^{\pi}\left(  z\right)  =z-r_{\pi}^{2}$ and
$a^{K}\left(  z\right)  =z-r_{\pi}^{2}-1$ at $\mathcal{O}\left(  p^{4}\right)
$). The resulting total rate, taking the $\pi^{+}\pi^{-}$ contribution alone
for simplicity, can be written as%
\begin{align*}
\Gamma_{\ell^{+}\ell^{-}} &  =\frac{G_{8}^{2}m_{K}^{5}\alpha^{4}}{512\pi^{7}%
}\int_{4r_{\ell}^{2}}^{\left(  1-r_{\pi}\right)  ^{2}}dz\;\left|  a\left(
z\right)  \right|  ^{2}\lambda_{\pi}^{1/2}\left|  \mathcal{J}\left(
\frac{r_{\pi}^{2}}{z},\frac{r_{\ell}^{2}}{z}\right)  \right|  ^{2}%
\frac{r_{\ell}^{2}}{z}\left(  1-\frac{4r_{\ell}^{2}}{z}\right)  ^{3/2}\\
\Gamma_{\gamma\gamma} &  =\frac{G_{8}^{2}m_{K}^{5}\alpha^{2}}{4096\pi^{5}}%
\int_{0}^{\left(  1-r_{\pi}\right)  ^{2}}dz\;\left|  a\left(  z\right)
\right|  ^{2}\lambda_{\pi}^{1/2}\left|  \mathcal{J}_{\gamma\gamma}\left(
\frac{r_{\pi}^{2}}{z}\right)  \right|  ^{2}%
\end{align*}
where $r_{x}=m_{x}/m_{K}$, $\lambda_{\pi}=\lambda\left(  1,r_{\pi}%
^{2},z\right)  $ is the standard two-body kinematical function for $\pi
^{0}+\left(  \pi^{+}\pi^{-}\right)  _{0^{++}}$ in a $L=0$ wave, $r_{\ell}%
^{2}/z$ is the helicity suppression factor and $\left(  1-4r_{\ell}%
^{2}/z\right)  ^{3/2}$ stands for the lepton pair in a $L=1$ wave. The
two-loop function $\mathcal{J}$ describes the transitions $\left(  \pi^{+}%
\pi^{-}\right)  _{0^{++}}\rightarrow\ell^{+}\ell^{-}$. Care is compulsory in
computing this function for varying $z$, because of compensating IR
divergences among the $\pi\pi$ and $\pi\pi\gamma$ cuts. Various numerical
tests were performed, in particular analyticity of $\mathcal{J}$ as a function
of $z$.

\begin{figure}[t]
\begin{center}%
\[
\includegraphics[width=13cm]{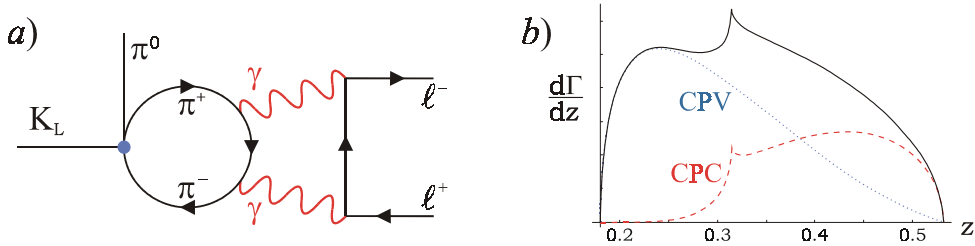}  \hspace{1cm}%
\]
\end{center}
\caption{(a) Typical CPC pion loop contribution and (b) the CPV and CPC
differential rates for positive interference between DCPV and ICPV.}%
\label{Fig2}%
\end{figure}

We have included the standard $K_{L}\rightarrow\pi^{0}\gamma\gamma$
parametrization (see \cite{CEP}) with the obvious intent of taking the ratio
$R_{\gamma\gamma}=\Gamma_{\ell^{+}\ell^{-}}/\Gamma_{\gamma\gamma}$. The
crucial point is that for a large range of parametrization of the vertex
$K_{L}\rightarrow\pi^{0}P^{+}P^{-}$, $R_{\gamma\gamma}$ is stable even if both
the $\ell^{+}\ell^{-}$ and $\gamma\gamma$ spectra vary much. For dynamical
reasons, the $e^{+}e^{-},\mu^{+}\mu^{-}$ and $\gamma\gamma$ modes react
similarly to modulations in the distribution of momentum entering the scalar
subprocess (i.e., to $a^{P}\left(  z\right)  $). Given this observation, we
infer the branching ratio of $\ell^{+}\ell^{-}$ modes from the experimental
result for $\gamma\gamma$. In doing so, some higher order chiral corrections
are included in our result, in particular the $\mathcal{O}\left(
p^{6}\right)  $ chiral counterterms (with their VMD contents) needed to
describe both the rate and spectrum for $K_{L}\rightarrow\pi^{0}\gamma\gamma$.
The stability of $R_{\gamma\gamma}$ is the key dynamical feature permitting
such an extrapolation, and thereby getting a reliable estimation for
$\Gamma_{\mu^{+}\mu^{-}}$.

Numerically, taking $B\left(  K_{L}\rightarrow\pi^{0}\gamma\gamma\right)
^{\exp}=\left(  1.42\pm0.13\right)  \times10^{-6}$ as the average of the
KTeV\cite{KTeV_KLpgg} and NA48\cite{NA48_KLpgg} measurements, we find
$B\left(  K_{L}\rightarrow\pi^{0}\mu^{+}\mu^{-}\right)  _{CPC}^{0^{++}%
}=\left(  5.2\pm1.6\right)  \times10^{-12}$, with a conservative error
estimate of 30\%.

Finally, the differential rate is in Fig.2b, and shows that an appropriate
kinematical cut can reduce the relative contamination of the CPC contribution
to below 10\%.

\section{Phenomenological Analysis}

The final parametrizations are, in the Standard Model ($\kappa=10^{4}%
\operatorname{Im}\lambda_{t}=1.36\pm0.12$)%
\[%
\begin{array}
[c]{rlr}%
B\left(  K_{L}\rightarrow\pi^{0}\nu\bar{\nu}\right)  \approx & \left(
16\kappa^{2}\right)  \times10^{-12} & \text{\cite{BSU}}\\
B\left(  K_{L}\rightarrow\pi^{0}e^{+}e^{-}\right)  \approx & \left(
2.4\kappa^{2}\pm6.2\left|  a_{S}\right|  \kappa+15.7\left|  a_{S}\right|
^{2}\right)  \times10^{-12} & \text{\cite{BDI}}\\
B\left(  K_{L}\rightarrow\pi^{0}\mu^{+}\mu^{-}\right)  \approx & \left(
1.0\kappa^{2}\pm1.6\left|  a_{S}\right|  \kappa+3.7\left|  a_{S}\right|
^{2}+5.2\right)  \times10^{-12} & \;\;\;\;\;\text{\cite{OurWork}}%
\end{array}
\]
The $\kappa^{2}$ terms are DCPV, the ICPV parameter $a_{S}$ is the counterterm
dominating $K_{S}\rightarrow\pi^{0}\ell^{+}\ell^{-}$, recently measured by
NA48/1 as $\left|  a_{S}^{\exp}\right|  =1.2\pm0.2$\cite{NA48as}. The
interference between DCPV and ICPV has been argued to lead to positive
sign\cite{BDI,FGD}.

It is interesting to keep track of the underlying short-distance physics,
especially for analyzing the possible impact of New Physics. The FCNC diagrams
in Fig.1a,b, including QCD corrections, leads to the effective Hamiltonian
\cite{OPE}%
\[
H_{eff}^{\left|  \Delta S\right|  =1}=\frac{G_{F}\alpha}{\sqrt{2}}V_{ts}%
^{\ast}V_{td}\left(  y_{7V}Q_{7V}+y_{7A}Q_{7A}\right)  \;\;\;\text{with}%
\;Q_{7V\left(  A\right)  }=\left(  \bar{s}d\right)  _{V-A}\otimes\left(
\bar{\ell}\ell\right)  _{V\left(  A\right)  }%
\]
with $y_{7V}=0.73\pm0.04$ (for $\mu\simeq1$ GeV) and $y_{7A}=-0.68\pm0.03$.
The various CPV coefficient dependences on these Wilson coefficients are (ICPV
is long-distance)%
\[%
\begin{array}
[c]{lllll}%
C_{ICPV}^{e}=15.7 &  & C_{int.}^{e}=8.91y_{7V} &  & C_{DCPV}^{e}=2.67\left(
y_{7A}^{2}+y_{7V}^{2}\right) \\
C_{ICPV}^{\mu}=3.7 &  & C_{int.}^{\mu}=2.12y_{7V} &  & C_{DCPV}^{\mu
}=0.63\left(  y_{7A}^{2}+y_{7V}^{2}\right)  +0.85y_{7A}^{2}%
\end{array}
\]
There is a simple phase-space suppression factor for all terms, but somewhat
compensating this, the muon mode DCPV receives an additional
helicity-suppressed axial-vector FCNC contribution. This gives different
sensitivity of the two modes to the underlying short-distance
physics.\begin{figure}[t]
\begin{center}%
\[
\includegraphics[width=11cm]{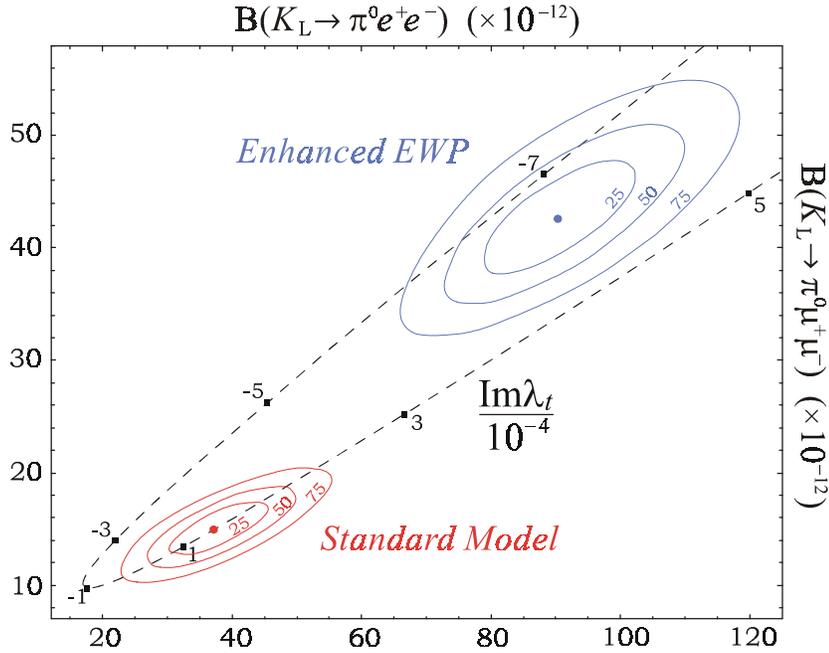}  \hspace{1cm}%
\]
\end{center}
\caption{The confidence regions for the SM and EEWP scenario.}%
\label{Fig3}%
\end{figure}

Plotting the muon mode against the electron one in terms of $\operatorname{Im}%
\lambda_{t}$ (see Fig.3), we get a curve whose spreading is directly related
to the relative strength of the vector and axial vector FCNC. This plane is
especially suited to study New Physics impacts. Let us take as an example the
enhanced electroweak penguin model (EEWP) of Buras \textit{et al}\cite{BFRS},
which affects the Wilson coefficients as $y_{7V}^{EEWP}\approx0.9$ and
$y_{7A}^{EEWP}\approx-3.2$. Taking all errors into account, we get the
theoretical predictions for positive interference as in Fig.3, with the
corresponding branching ratios%
\[%
\begin{tabular}
[c]{llll}
& \ \ S.M. $\left(  \times10^{-11}\right)  \;\;$ & \ \ EEWP $\left(
\times10^{-11}\right)  \;$\cite{BFRS}$\;$ & \ \ \ Experiment \cite{KTeV}%
\ \ \ \\\hline
$K_{L}\rightarrow\pi^{0}\nu\bar{\nu}$\cite{BSU} & \multicolumn{1}{c}{$3.0\pm
0.6$} & \multicolumn{1}{c}{$31\pm10$} & \multicolumn{1}{c}{$<5.9\times10^{-7}%
$}\\
$K_{L}\rightarrow\pi^{0}e^{+}e^{-}$ & \multicolumn{1}{c}{$3.7_{-0.9}^{+1.1}$}
& \multicolumn{1}{c}{$9.0\pm1.6$} & \multicolumn{1}{c}{$<2.8\times10^{-10}$}\\
$K_{L}\rightarrow\pi^{0}\mu^{+}\mu^{-}$ & \multicolumn{1}{c}{$1.5\pm0.3$} &
\multicolumn{1}{c}{$4.3\pm0.7$} & \multicolumn{1}{c}{$<3.8\times10^{-10}$%
}\\\hline
\end{tabular}
\ \ \ \ \
\]
Note that the central value for EEWP is not on the $\operatorname{Im}%
\lambda_{t}$ curve as the ratio $(y_{7V}/y_{7A})^{SM}\neq(y_{7V}%
/y_{7A})^{EEWP}$.

\ \newline 

In conclusion, the set of rare $K_{L}$ decay modes $\pi^{0}\nu\bar{\nu}$,
$\pi^{0}e^{+}e^{-}$ and $\pi^{0}\mu^{+}\mu^{-}$ is now fully under theoretical
control, and provides for a stringent test of the Standard Model. In addition,
if a signal of New Physics is found, information on its nature can be
extracted from the observed pattern of branching ratios.

Experimentally, it is clear that the $K_{L}\rightarrow\pi^{0}\mu^{+}\mu^{-}$
mode should be seriously considered. Also, additional measurements of
$K_{S}\rightarrow\pi^{0}\ell^{+}\ell^{-}$ would be certainly very desirable
since the main uncertainty on the theoretical prediction for the CPV parts,
and thus the spreading of the confidence regions in Fig.3, comes from $a_{S}$.

\section{Acknowledgements}

This work has been supported by IHP-RTN, EC contract No. HPRN-CT-2002-00311 (EURIDICE).

\end{document}